\documentclass[prb,twocolumn,superscriptaddress,showpacs]{revtex4-1}

\usepackage{graphicx,bm,color}

\usepackage{graphicx}
\usepackage{dcolumn}
\usepackage{bm}
\usepackage{mhchem}
\usepackage{epstopdf}
\usepackage{color}

\begin{document}


\title{Magnetic Exchange Interaction in the Spin Polarized Electron Gas }

\author{Mohammad M. Valizadeh}
\email{mvbr5@mail.missouri.edu}

\affiliation{Department of Physics $\&$ Astronomy, University of Missouri, Columbia, MO 65211, USA}

\author{S. Satpathy}

\affiliation{Department of Physics $\&$ Astronomy, University of Missouri, Columbia, MO 65211, USA}



\date{\today}

\begin{abstract}

The exchange interaction between two magnetic moments embedded in a host metal is fundamental to the description of the magnetic behavior of solids. In the standard spin-degenerate electron gas, it leads to the well known
Ruderman-Kittel-Kasuya-Yoshida (RKKY) interaction, which is of the Heisenberg form $J\vec S_1\cdot   \vec S_2$, where the interaction strength $J$ oscillates and falls off with the distance between the moments in a characteristic way. Here, we study the more general case of the spin-polarized electron gas both in two and three dimensions, by evaluating the interaction strength as an integration over the product of the host Green's functions. We find that in addition to the Heisenberg term, an additional Ising-like term appears in the magnetic interaction, so that the net interaction for the spin-polarized gas is of the form  $J_1 \vec S_1\cdot   \vec S_2 + J_2 S_{1z}   S_{2z}$. The interactions show  a beating pattern as a function of distance, caused by the two different Fermi momenta for the two spins.

\end{abstract}

\maketitle 

\section{Introduction}

The Ruderman-Kittel-Kasuya-Yosida (RKKY) interaction~\cite{RKKY1,RKKY2,RKKY3} is an indirect exchnage interaction between two localized magnetic moments, 
mediated by electrons of the host crystal, and it has been extensively studied in one ~\cite{RKKY-1D}, two ~\cite{RKKY-2D}, or three dimensions ~\cite{RKKY-3D}. 
For a system with no broken symmetry (inversion or time reversal), this interaction has the Heisenberg form of $E(\vec R) = J  \vec S_1 \cdot \vec S_2$.  For the standard free electron gas, $J$ shows an oscillatory behavior
as a function of the distance $\vec R$ between the two moments, with the long-distance behavior $J(\vec R)\sim \cos(2k_F R)/R^{d}$, where $k_F$ is the Fermi momentum and $d$ is the dimensionality of the system.
The recent observation of the skyrmions in solids~\cite{skyrmea1962,Tokura,sk1,sk2}, caused due to the competition between RKKY and Dzyaloshinsky-Moriya (DM) interactions, originating from broken symmetry and spin-orbit interaction,
has stimulated considerable interest in systems with broken symmetry.

The spin polarized electron gas, which we consider here, is the simplest example of a system with broken symmetry,
and serves to illustrate the effect of the broken symmetry on the magnetic interaction.
Time reversal and inversion symmetries lead, respectively, to the conditions for the band structure energies: 
$\varepsilon_{\vec k\uparrow}=\varepsilon_{-\vec k\downarrow}$ and 
$\varepsilon_{\vec k\uparrow}=\varepsilon_{-\vec k\uparrow}$.
These conditions together lead to the spin-degenerate energies in the band structure, viz., 
$\varepsilon_{\vec k\uparrow}=\varepsilon_{\vec k\downarrow}$. 
Note that the spin-polarized electron gas has the inversion symmetry present, but the time reversal symmetry is broken.

Broken symmetries, break this degeneracy of the energy bands, change the Green's function matrix, and finally lead to extra terms in the interaction between two localized magnetic moments.
Using lattice models, Dzyaloshinski and Moriya  showed that~\cite{DM,Moriya} in certain situations with broken inversion symmetry,
the net interaction is given by the expression:
$E(\vec R)=J(\vec R)\vec S_1\cdot   \vec S_2+\vec D(\vec R)\cdot\vec S_1\times\vec S_2+\vec S_1 \cdot  \stackrel{\leftrightarrow}{\Gamma}  \cdot  \vec S_2$,
where in addition to the scalar RKKY-type interaction, we also have the vector and tensor interactions between the two localized moments
 $\vec S_1$ and $\vec S_2$. For the spin-polarized electron gas discussed below, we will find that the scalar and tensor terms are non zero and their magnitudes are such that the net interaction may be written as
 \begin{equation}
 E(R)=J_1 \vec S_1\cdot   \vec S_2 + J_2 S_{1z}   S_{2z}. 
 \end{equation}
 The DM vector interaction $\vec D$ turns out to be zero, because the present system is inversion symmetric.
 
\section{Expression for magnetic interaction}
 
 As usual, we take the localized moments to interact with the host electrons, described by 
 the Hamiltonian $\hat H$, via the contact interaction
\begin{equation}
 V_1 (\vec r)=-\lambda \,\delta(\vec r)\,\vec S_1\cdot \vec s,
\label{eq2}
\end{equation}
and
\begin{equation}
V_2 (\vec r)=-\lambda \,\delta(\vec r-\vec R)\,\vec S_2\cdot \vec s,
\end{equation}
where $\vec s$ is the spin of the electron.
Using the second-order perturbation theory, one can evaluate the interaction energy between the two localized spins.
The result is
\begin{align}
E(\vec R)= & \frac{-\lambda^2}{\pi} \ {\rm Im} \int_{-\infty}^{E_F} Tr \big[G(0,\vec R, E)\, \vec S_2\cdot\vec s \nonumber \\
           & G(\vec R, 0, E) \vec S_1\cdot\vec s\, \big]dE,
\label{master}
\end{align}
where 
$\hat G(E)=(E+i\mu-\hat H)^{-1}$ with $\mu\rightarrow 0^+$,  is the retarded Green's function. 
The matrix elements are given by
\begin{equation}
G_{\sigma_1\sigma_2}(\vec r_1,\vec r_2,E)=\sum \limits_{\vec k\nu}^{\infty}  \frac{\psi_{\vec k\nu}  (\vec r_1,\sigma_1)\psi_{\vec k\nu}^*(\vec r_2,\sigma_2)}{E+i\mu-\varepsilon_{\vec k\nu}},
\label{GF}
\end{equation}
where $G_{\sigma_1\sigma_2}(\vec r_1,\vec r_2,E) \equiv \langle \vec r_1 \sigma_1 | \hat G (E)| \vec r_2 \sigma_2 \rangle$,
$\psi_{\vec k\nu}(\vec r,\sigma)  =    \langle\vec r\sigma |   \vec k\nu\rangle$,
and $\vec k\nu$ labels the eigenstates of the system.

The eigenstates are in general spin mixed, but in the present case they are spin pure states, so that the Green's function
is diagonal in the spin indices. 
Furthermore, if the wave functions can be chosen to be real (true if $\psi_{\vec k\nu}  (\vec r,\sigma)$ and $\psi^*_{\vec k\nu}  (\vec r,\sigma)$ are solutions with the same energy), then it follows from Eq. (\ref{GF}) that 
$G_{\sigma_1\sigma_2}(\vec r_1,\vec r_2,E)  =   G_{\sigma_2\sigma_1}(\vec r_2,\vec r_1,E)$. In the present case, the Green's function being spin diagonal, we have  the equality
 $G(0,\vec R, E)= G(\vec R, 0, E)$.
 
Under these conditions, we can expand the Green's function matrix  in terms of $\vec \sigma$, the Pauli matrices as
\begin{equation}
G(0,\vec R, E)= G(\vec R, 0, E) = g_0 (E)  \sigma_0 + g_z (E) \sigma_z,
\label{g0gz}
\end{equation}
$\sigma_0$ being the unit $2 \times 2$ matrix. The energy expression Eq. \ref{master} can be evaluated using the following spin identities
\begin{align}
& {\rm Tr }   \big [\vec S_2\cdot \vec \sigma \  \vec S_1 \cdot  \vec \sigma\big ]=2\vec S_1\cdot\vec S_2,   \nonumber\\
& {\rm Tr }   \big [\vec S_2  \cdot \vec \sigma \   \sigma_z  \  \vec S_1\cdot\vec \sigma\big ] = 2 i(\vec S_1\times\vec S_2)_{z},   \nonumber\\
& {\rm Tr }   \big [ \sigma_z \ \vec S_2  \cdot \vec \sigma \   \sigma_z  \  \vec S_1\cdot\vec \sigma\big ] = 
-2\vec S_1\cdot\vec S_2    +4 S_{1z} S_{2z}.
\label{identities}
\end{align}
The result is
\begin{equation}
E(\vec R) =   J_1 \vec S_1 \cdot \vec S_2 + J_2 S_{1z}S_{2z},
\end{equation}
where
\begin{align}
& J_1 = \frac{-\lambda^2 \hbar^2}{2 \pi} \times  {\rm Im}  \int_{-\infty}^{E_F}  (g_0^2 - g_z^2) \ dE,   \nonumber\\
&J_2 = \frac{-\lambda^2 \hbar^2}{2 \pi} \times  {\rm Im}    \int_{-\infty}^{E_F} 2 g_z^2 \  dE.
\label{J1J2}
\end{align}
These expressions can be evaluated from the  Green's functions, which we now proceed to do for the spin-polarized electron gas in 2D and 3D. 
Note from Eq. \ref{J1J2} that for the standard  (spin  unpolarized) electron gas, the Green's function 
has equal diagonal elements, so that $g_z = 0$, and the $J_2$ terms vanishes as a result and one obtains the standard $\vec S_1 \cdot \vec S_2$ RKKY interaction.


\section{Spin polarized electron gas in 3D}

Our starting point is the electron band structure
\begin{equation}
\varepsilon_{\vec k \sigma}    =  \frac{\hbar^2 k^2}{2m} \mp \Delta,
\label{3DE}
\end{equation}
where  $- (+)$ sign is for spin up (down) states, so that $2 \Delta$ is the band splitting between the up- and down-spin states. 
The corresponding plane-wave eigenstates are
\begin{equation}
| \vec k \sigma    \rangle = \frac{1}{\sqrt \Omega} e^{i \vec k \cdot \vec r} |\sigma \rangle,
\label{psi}
\end{equation}
where the $\Omega$ is the volume of the box for normalization.
The key quantity to evaluate is the Green's function, which, using Eqs. (\ref{GF}), (\ref{3DE}), and (\ref{psi}) and converting the summation into integration in the momentum space, is written as  
\begin{align}
 G_{\sigma\sigma^\prime}(\vec r,\vec r \ ^\prime,E)   
=\frac{\delta_{\sigma\sigma^\prime}}      {(2 \pi)^3}   
\int 
\frac{e^{i\vec k\cdot(\vec r-\vec r \  ^\prime)}}   {E+i\eta-\varepsilon_{\vec k \sigma}}  \ d^3k.
\end{align}
The integral 
 can be evaluated by a straightforward
contour integration~\cite{RKKY-3D,Valizadeh}   to yield the result 
\begin{align}
&  G(\vec r_1, \vec r_2,E)
 = \begin{pmatrix}
g_{\uparrow} & 0 \\
0 & g_{\downarrow} 
\end{pmatrix} \nonumber \\
& = \frac{-m}{2\pi r\hbar^{2}}
\begin{pmatrix}
e^{i\alpha(E+\Delta) r} & 0 \\
0 & e^{i\alpha(E-\Delta)r} 
\end{pmatrix},
\label{GF3D}
\end{align}
where $r \equiv | \vec r_1 - \vec r_2|$ and
\begin{eqnarray}
\alpha (x) =\begin{cases}
 (2m\hbar^{-2} x)^{1/2}  \text { if  $ x > 0$},   \\
i  (2m\hbar^{-2}|x|)^{1/2}  \text { if  $ x < 0$}.
\end{cases}
\end{eqnarray}
The coefficients $g_0$ and $g_z$  in Eq. (\ref{g0gz})   are then $g_0  = (  g_{\uparrow} + g_{\downarrow})/2$ and 
$g_z  = (  g_{\uparrow} -  g_{\downarrow})/2$, which are complex numbers. 

Plugging these into Eq. (\ref{J1J2}) and performing the 
energy integrations, we find that the imaginary part vanishes, as it must, and the results for the magnetic interaction terms are given by
\begin{align}
& J_1=\frac{-\lambda^2\,m^2}{8\pi^3\hbar^2\,R^2}\      \times   \bigg (      \int_{\Delta}^{E_F} \sin  \big [ k_+ (E) R + k_- (E) R)    \big]
                                             \       dE     \nonumber \\
                                   &           \hspace{20mm}   +  \int_{-\Delta}^{\Delta} \      \exp [-\kappa (E) R] \times 
                                                 \sin  \big [ k_+ (E) R    \big]  \  dE      \bigg ),     \nonumber \\
                                             & J_2 =      I   (     k_{F-} R)+ I   (      k_{F+}R)   -J_1,
\label{J1}	
\end{align}
where $E_F$ is the Fermi energy,
$ k_\pm (E) = [   2m\hbar^{-2} (E \pm \Delta) ]^{1/2}$ is the momentum for the spin up (down) state, 
$k_{F\pm} \equiv k_\pm (E_F)$ is the corresponding Fermi momentum for spin up (down) electrons, 
$ \kappa (E) = [   2m\hbar^{-2} (\Delta - E) ]^{1/2}$, and 
\begin{equation}
I(x)=-\frac{\lambda^2m}{(4\pi)^3R^4}     \times  [ \sin(2x)-2x\,\cos(2x\,) ].    	
\end{equation}
The integral $I(x)$ is familiar from the theory of the spin-unpolarized electron gas, in which case, the RKKY interaction is, simply,  
\begin{equation}
J_1 = 2 I(k_F R), 
\label{J13D}  	
\end{equation}
 $k_F =  (2m\hbar^{-2} E_F)^{1/2} $ being the Fermi momentum. The results agree with our earlier work, where we had used a different method \cite{Valizadeh}.

\begin{figure}[htp]%
\includegraphics[scale=0.37]{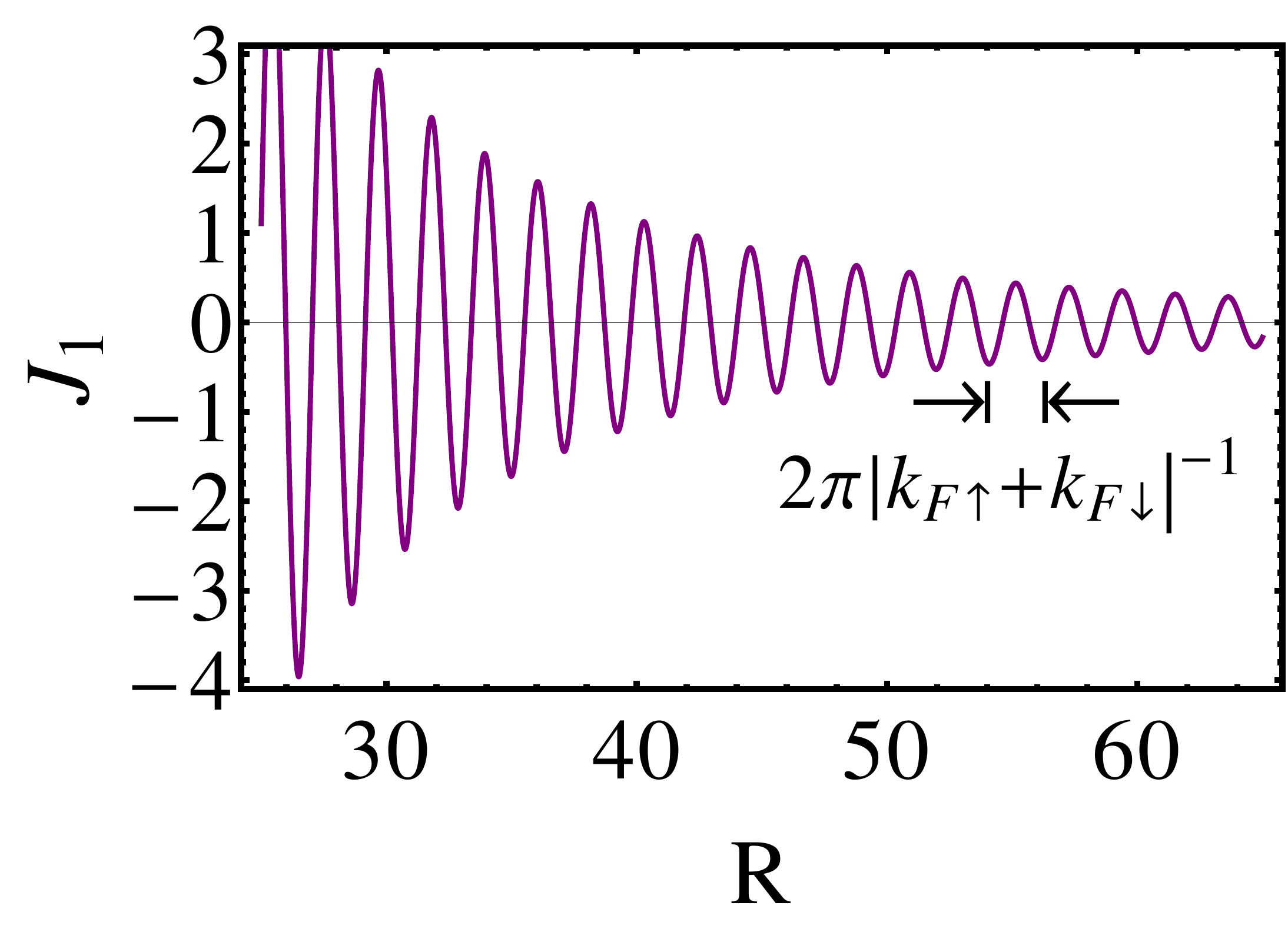}
\includegraphics[scale=0.37]{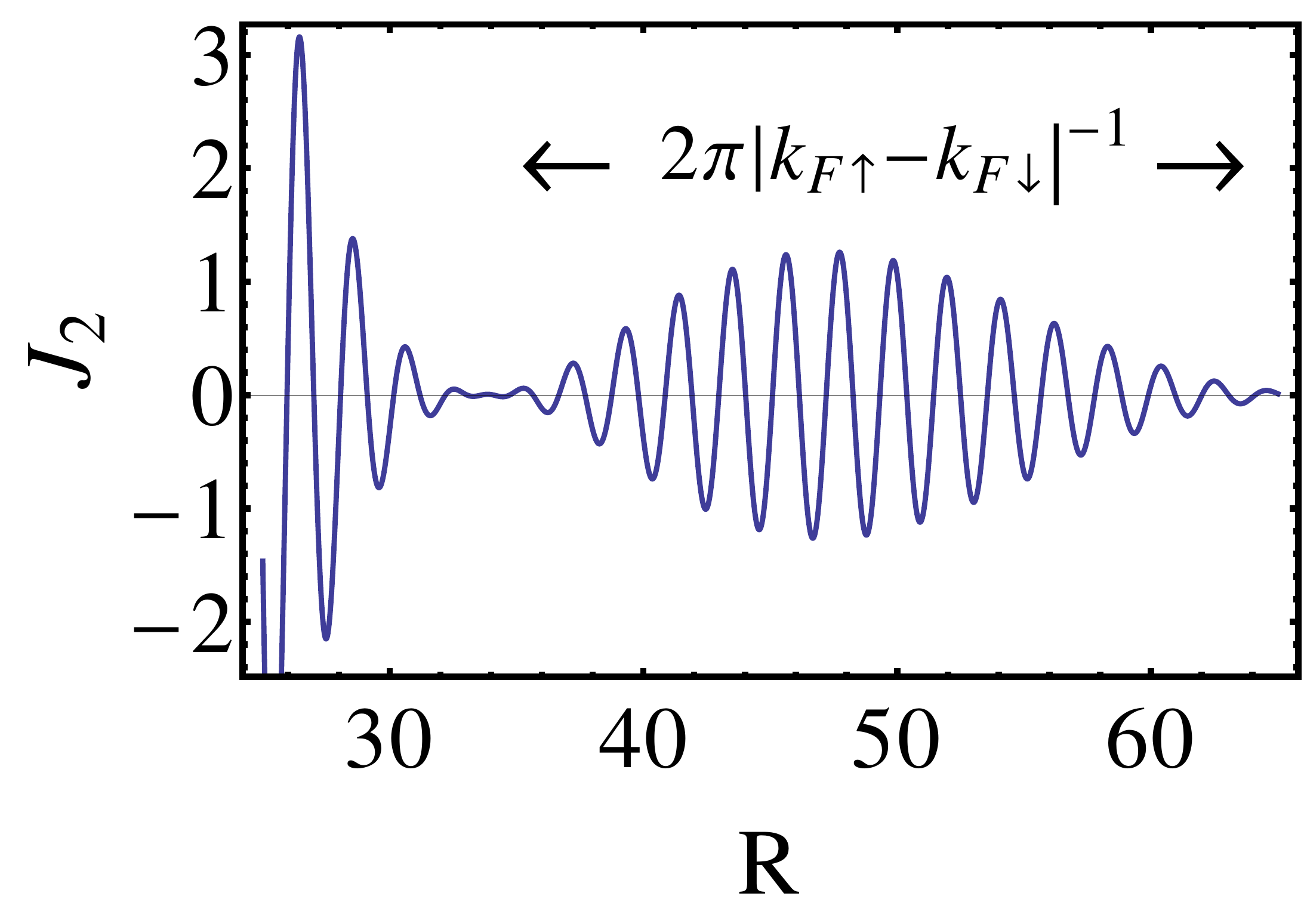}
\caption{%
Oscillatory behavior of $J_1$ and $J_2$, in the units of $10^{-5}m \lambda^{2}\pi^{-3}$ as a function of $R$ in the units of \AA\ for the case of iron.
}
\label{fig2}
\end{figure}

Note that if there is no spin polarization ($\Delta = 0$), then Eq. (\ref{J1}) immediately reduces 
to the expression Eq. (\ref{J13D}) for the spin-unpolarized gas and $J_2 = 0$. 
The computed results for $J_1$ and $J_2$, for the case of iron, are shown in Fig. \ref{fig2}, which show the oscillatory behavior characteristic of the inverse momentum $(k_{F+} + k_{F-})^{-1}$, and the beat pattern for $J_2$ is characteristic of the inverse difference $ (k_{F+} - k_{F-})^{-1}$, respectively.
Band calculations for iron \cite{Callaway} lead to the $E_F \approx 8 $ eV and $2 \Delta \approx 2 $ eV. 
One can use these values to find the Fermi momenta, $k_{F\uparrow}\simeq1.574$ and $k_{F\downarrow}\simeq1.388$ 1/\AA\ .
For long distances, we predict the oscillation periods to be: $2 \pi |k_{F\uparrow} + k_{F\downarrow}|^{-1}\simeq2.12$ \AA\, and for the beat pattern behavior $2 \pi |k_{F\uparrow} - k_{F\downarrow}|^{-1}\simeq 33.75 $ \AA\ . 
Fig. \ref{fig2} shows a very good match between the computed results and predictions.

{\it Discussions} -- An interesting situation occurs if $J_1 + J_2 = 0$, which can happen for certain distances. In this case, the net interaction, has the form of
$E(\vec R) = J_1 ( S_{1x} S_{2x} + S_{1y} S_{2y})$, which would clearly align the spins in the $xy$-plane, i. e., normal to the spin polarization axis. In general, depending on the relative strengths of $J_1 (\vec R) $ and $J_2 (\vec R)$, the net spin interaction could align the two spins in different directions, leading to the possibility for unusual spin textures.

%

\begin{figure}[h]%
\centering
\includegraphics[scale=0.37]{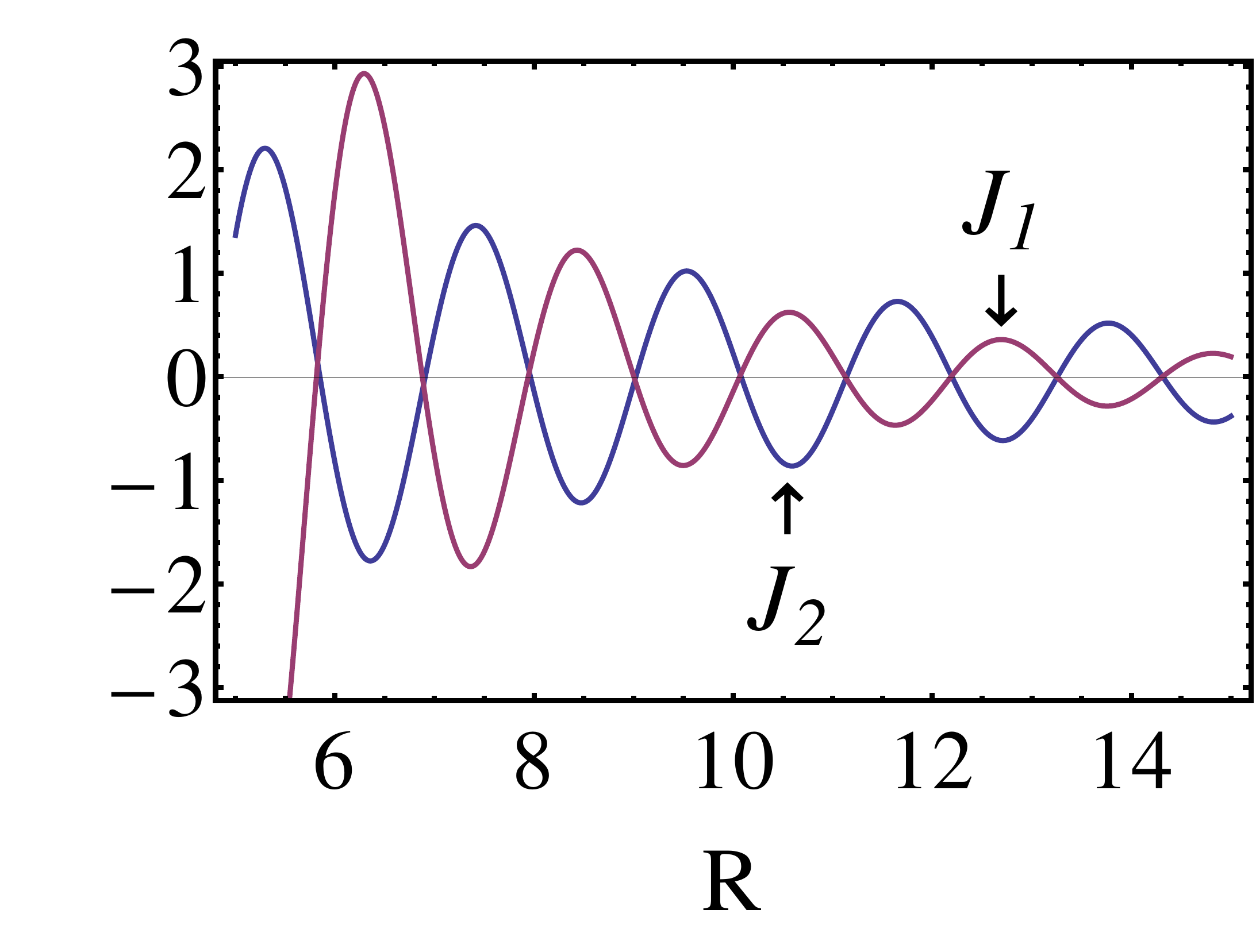}
\caption{Oscillatory behavior of $J_1$ and $J_2$, in the units of $10^{-3}m \lambda^{2}\pi^{-3}$ as a function of $R$ in the units of \AA\ for the case of iron.}
\label{fig3}
\end{figure}

Note that apart from the position dependent interaction $E (\vec R)$, there is a constant energy shift
\begin{equation}
E_0=\frac{-\hbar\lambda}{12\pi^2}(S_{1z}+ S_{2z})(k_{F+}^3-k_{F-}^3), 
\label{E0-3D}
\end{equation}
a new term not present in the standard, non-spin-polarized electron gas, and a term that tends to align the embedded spins $\vec S_1$
and $\vec S_2$ along the $\hat z$ axis. This expression\cite{Valizadeh}, obtained by using the first-order perturbation theory for the perturbing potentials, Eqs. (\ref{eq2}) and (\ref{eq3}), depends on the strength of the polarization of the electron gas and would dominate for strong spin polarization. In turn,
for weak spin polarization 
($\delta k_F \equiv k_{F+} - k_{F-} \ll  m \lambda \bar k_F^2 /(R \bar k_F)^4$, 
where $\bar k_F \equiv (k_{F+} - k_{F-})/2$), 
this term is negligible and the $J_1$ and $J_2$ interactions dominate.

Another point to note is that the essential ingredient for the presence of the DM interactions 
 is the broken symmetry (time reversal or inversion or both). In the original DM work\cite{Moriya}, the spin-orbit coupling (SOC)  provided the mechanism for the magnetic interaction. 
 The interaction between the magnetic moments of two atoms occurred via an intermediate atom and involved the spin-orbit coupled excited states on the two atoms. In this case, as originally showed by DM, $J \sim \xi^0$, $\vec D \sim \xi$,
 and $\stackrel{\leftrightarrow}{\Gamma} \sim \xi^2$, where $\xi$ is the spin-orbit coupling strength
 ($\xi  \ \vec L \cdot \vec S$), so that  $| \stackrel{\leftrightarrow}{\Gamma}| \ll | \vec D | \ll |J|$, $\xi$ being a small parameter, and it is then customary to ignore the tensor DM interaction $\stackrel{\leftrightarrow}{\Gamma}$. 
 In the present case, the broken time-reversal symmetry without the involvement of any SOC leads to the DM interaction, so that it is entirely a different mechanism, and further that the strengths of all terms are comparable, being proportional to $\lambda ^2$. 
Thus, in a solid, if a spin-polarized electron gas is present in addition to magnetic moments on atoms with SOC, both effects must be considered separately and the dominant effect for the DM interactions might as well come from the spin polarization of the electron gas, of the type studied in this paper.

\section{Spin polarized electron gas in 2D}

The Green's function for the spin polarized electron gas in 2D is given by 
\begin{align}
 G_{\sigma\sigma^\prime}(\vec r,\vec r \ ^\prime,E)   
=\frac{\delta_{\sigma\sigma^\prime}}      {(2 \pi)^2}   
\int 
\frac{e^{i\vec k\cdot(\vec r-\vec r \  ^\prime)}}   {E+i\eta-\varepsilon_{\vec k \sigma}}  \ d^2k.
\end{align}
This a standard integration, which can be evaluated by using the Jacobi-Anger expansion of the exponential term in terms of the 
Bessel's functions and by performing the angular integration.\cite{Jacobi-Anger,Litvinov,Valizadeh}  The result is 
\begin{equation}
G_{\sigma \sigma } (\vec r,\vec r \ ^\prime,E)  = -\frac{m}{\pi \hbar^2}K_0 \left[ -i  \frac{\sqrt {2m} }{\hbar}   |\vec r - \vec r \ ^\prime|
\   \alpha (E \pm \Delta)   \right],
\label{eq15}
\end{equation}
where + (-) is for $\sigma = \uparrow$ ($\downarrow$), $K_0$ is the modified Bessel function of the second kind, and
\begin{eqnarray}
\alpha (x) =\begin{cases}
\sqrt x & \text{if $ x >  0$},\\
i \sqrt {|x |} & \text{if $ x <  0$}.
\end{cases}
\label{eq16}
\end{eqnarray}
To find the imaginary and the real parts of the modified Bessel function, it is convenient to
use the equality $K_{\nu}(x)=   2^{-1} \pi \  i^{\nu+1}H_{\nu}^{1}(ix)$, which is valid for $-\pi<   \arg \ (x)\leq  \pi  / 2$. 
The Hankel function of the first kind is written in terms of the 
Bessel and Neumann functions as $H{_\nu}^{1}(x) = J_\nu (x) + i Y_\nu (x)$. 
The expansion coefficients for the Green's function, Eq. (\ref{g0gz}), are then $g_0 = 2^{-1} (G_{\uparrow \uparrow} (\vec r,\vec r \ ^\prime,E) + G_{\downarrow \downarrow} (\vec r,\vec r \ ^\prime,E) )$ and 
$g_z = 2^{-1} (G_{\uparrow \uparrow} (\vec r,\vec r \ ^\prime,E) - G_{\downarrow \downarrow} (\vec r,\vec r \ ^\prime,E) )$.
Plugging these into Eq. (\ref{J1J2}), we find the results
\begin{align}
  J_1   &=\frac{\lambda^2\,m^2}{8\pi\hbar^2} \   \Big\{ 
          -\frac{2}{\pi}  \int_{-\Delta}^{\Delta}Re\big     [ K_0 (  \kappa R ) ]
          J_0 (    k_+ R)  dE     \nonumber \\
          + &\int_{\Delta}^{E_F}    [J_0 ( k_-R)  \    Y_0   (k_+R) 
          + Y_0 (k_-R) \    J_0 (k_+R) ]   \  dE   \Big     \},    \nonumber   \\
           J_2   & =    \frac{\lambda^2m}{16\pi R^2}         [  I^\prime (k_{F-} R ) +  I^\prime ( k_{F+} R)  ]   -J_1,
\end{align}
where %
$   I^\prime(x)=   x^2    [   J_0(x)\,Y_0(x)+J_1(x)\,Y_1(x) ]$. 

In this case, similar to the case of 3D spin-polarized electron gas,
the oscillatory behaviors of $J_1$ and $J_2$ show beat-pattern, caused by the two different Fermi momenta for the two spin channels.

\section{Summary}

In this paper, we obtained the magnetic interactions between two localized moments, embedded in the 
 spin polarized electron gas in two and three dimensions, extending the standard results for the spin-unpolarized electron gas, which leads to the well known RKKY interaction. The spin-polarization leads to an anisotropic Heisenberg type of interaction, of the form $J_1 \vec S_1\cdot   \vec S_2 + J_2 S_{1z}   S_{2z}$. Both terms $J_1$ and $J_2$ show oscillatory behavior as a function of distance between the two   magnetic moments with the period of the oscillations determined by $\bar k_F R$, $\bar k_F $ being the average Fermi momentum of the two spin channels and, in addition, $J_2$ shows a beating pattern determined by the momentum difference $k_{F+} - k_{F-}$. This is the simplest system with broken symmetry and serves to illustrate the origin of the magnetic interactions in the solid that go  beyond the standard RKKY $\vec S_1\cdot   \vec S_2$ type interaction. 

\section{Acknowledgments}

This research was supported by the U.S. Department of Energy, 
Office of Basic Energy Sciences, Division of Materials Sciences and Engineering under Award  No. DE-FG02-00ER45818.  

\section{Appendix}

In this Appendix, we derive the general expression for the magnetic interaction, Eq. (\ref{master}), from the second order perturbation theory, which is somewhat more pedagogical than found in the literature.

Let $\psi_{\vec k\nu} (\vec r\sigma)$ denote the host electron wave functions
($|\vec k\nu \rangle = \sum_{\vec r\sigma} \psi_{\vec k\nu} (\vec r\sigma) |\vec r\sigma \rangle$ in Dirac notations), 
where $k\nu$ are the quantum numbers (e. g., Bloch momentum $k$ and band index $\nu$ in a crystal) and $\varepsilon_{k\nu}$ be the corresponding eigenenergies. The interaction between the two localized moments $\vec S_1$ and $\vec S_2$ (located at  origin and  $\vec R$, respectively) and the host electrons are taken, as usual, to be the contact interactions:
$V_1 (\vec r\,)=-\lambda \,\delta(\vec r\,)\,\vec S_1\cdot \vec s$
and
$
V_2 (\vec r\,)=-\lambda \,\delta(\vec r-\vec R)\,\vec S_2\cdot \vec s$. According to the second-order perturbation theory, the
change of energy due to this interaction is given by the equation
\begin{equation}
E(\vec R) = \sum_{\vec k\nu}^ {occ}     {\sum_{\vec {k^\prime} \nu^\prime} }^\prime \frac{| \langle \vec k \nu | \hat V_1 + \hat V_2 | \vec {k^\prime} \nu^\prime \rangle | ^2}
 {\varepsilon_{\vec k\nu} - \varepsilon_{\vec {k^\prime} \nu^\prime} },
 \label{ER}
\end{equation}
where the prime over the summation indicates that the term $k^\prime \nu^\prime = k \nu$ is excluded and the interactions are in the operator forms, viz.,  $\hat V_1 (\vec r)=-\lambda     \sum_{\sigma \sigma ^\prime}   |\vec 0 \sigma \rangle  \vec S_1 \cdot \vec s \     \langle \vec 0 \sigma^\prime |$
 and
  $\hat V_2 (\vec r)=-\lambda     \sum_{\sigma \sigma ^\prime}   |\vec R \sigma \rangle  \vec S_2 \cdot \vec s \     \langle \vec R \sigma^\prime |$. It is convenient to write the energy expression Eq. (\ref{ER}) in terms of the retarded and advanced Green's functions,  $\hat G(E)=(E+i\mu-\hat H)^{-1}$ and $\hat G^A(E)=(E - i\mu-\hat H)^{-1}$, where $\mu\rightarrow 0^+$. With the use of the identity
 \begin{equation}
 P (\frac{1}{x} ) =    \lim_{\mu \rightarrow \ 0 ^+ }
 \frac{1}{2} \big ( \frac {1}{x + i \mu} + \frac {1}{x - i \mu} \big ), 
 \end{equation}
 where $P$ denotes the principal part,
 and the expression for the Green's function
 \begin{equation}
\hat G (E) = \sum_{\vec k\nu} \frac {|\vec k \nu \rangle \langle \vec k \nu | } 
{E - \varepsilon_{\vec k \nu}+ i \mu},
 \end{equation}
 one finds after some algebra, the result
 \begin{equation}
E(\vec R)=\sum_{\vec k\nu}^{occ}   \langle\vec k\nu\mid\hat V_2 \   \hat G (\varepsilon_{\vec k\nu}) \     \hat V_1\mid\vec k\nu\rangle + h. c.,	
\label{ER2}
 \end{equation}
 where the extra terms ($\vec {k^\prime} \nu^\prime = \vec k \nu$)  added to Eq. (\ref{ER}) to write in terms of the Green's functions add up to zero. Note that only the cross terms in the interactions $\hat V_1$ and $\hat V_2$ have been kept, since only these
 depend on  $\vec R$. Using the completeness relation $\sum_{\vec r\sigma} |\vec r\sigma \rangle \langle \vec r\sigma| = 1$,
  Eq. (\ref{ER2}) can be expressed in terms of the real space wave functions
 \begin{align}
E(\vec R)= &   \lambda^2\sum_{\vec k\nu}^{occ} \sum\limits_{\sigma_1\sigma_2}
\langle\sigma_1\mid\vec S_2\cdot\vec s \,\, G(\vec R,0,\varepsilon_{\vec k\nu})\, \vec S_1\cdot\vec s  \mid \sigma_2\rangle \nonumber \\
& \times     \psi_{\vec k\nu}^*(\vec R,\sigma_1)\psi_{\vec k\nu}(0,\sigma_2) + h. c.,
\label{ER3}
\end{align}
where $G_{\sigma_1\sigma_2}(\vec r_1,\vec r_2,E) \equiv \langle \vec r_1 \sigma_1 | \hat G (E)| \vec r_2 \sigma_2 \rangle$
is given by
\begin{equation}
G_{\sigma\sigma'}(\vec r_1,\vec r_2,E)=\sum\limits_{\vec k\nu}   \frac{\psi_{\vec k\nu}(\vec r_1,\sigma)\psi_{\vec k\nu}^*(\vec r_2,\sigma')}{E+i\mu-\varepsilon_{\vec k\nu}}.
\end{equation}

Expressing the Green's function  as an integral over energy
\begin{equation}
G(\vec R,0,\varepsilon_{\vec k\nu})=\int G(\vec R,0,E)\,\delta(E-\varepsilon_{\vec k\nu}) \ dE,
\label{eq7}
\end{equation}
and the fact that $\int_{-\infty}^{\infty}   \  dE  \times   \sum_{\vec k\nu}   \rightarrow    \int_{-\infty}^{E_F}    \ dE  \times \sum_{\vec k\nu}^{occ}$,
Eq. (\ref{ER3}) leads to the result
 \begin{align}
E(\vec R)= &\lambda^2\int_{-\infty}^{E_F}dE \   \sum\limits_{\sigma_1\sigma_2}
\langle\sigma_1\mid\vec S_2\cdot\vec s \,\, G(\vec R,0,E)\, \vec S_1\cdot\vec s\mid\sigma_2\rangle
\nonumber \\
&  \times \sum_{\vec k\nu}    \psi_{\vec k\nu}^*(\vec R,\sigma_1)\psi_{\vec k\nu}    (0,\sigma_2)\delta(E-\varepsilon_{\vec k\nu})
+h.c.
\label{ER4}
\end{align}
The second line can be expressed as the difference between the retarded and the advanced Green's function, viz., 
 \begin{align}
\sum_{\vec k\nu}    \psi_{\vec k\nu}^*(\vec R,\sigma_1)\psi_{\vec k\nu}    (0,\sigma_2)\delta(E-\varepsilon_{\vec k\nu})
\nonumber \\
= \frac {i} {2 \pi}  [G(0, \vec R, E)  - G^A (0, \vec R, E)]_{\sigma_2 \sigma_1},
\label{GG}
\end{align}
since
 $\lim_{\mu\to 0^+}(x \pm i  \mu)^{-1}=P(x^{-1}) \mp   i\pi\delta(x)$, so that
\begin{equation}
\delta(E-\varepsilon_{\vec k\nu})=\frac{i}{2\pi}\Big(\frac{1}{E+i\mu-\varepsilon_{\vec k\upsilon}}-\frac{1}{E-i\mu-\varepsilon_{\vec k\upsilon}}\Big).
\end{equation}
From Eqs. (\ref{ER4}) and (\ref{GG}), we find the final result
\begin{align}
E(\vec R)= & \frac{-\lambda^2}{\pi} \ {\rm Im} \int_{-\infty}^{E_F} {\rm Tr} \big[G(0,\vec R, E)\, \vec S_2\cdot\vec s \nonumber \\
& G(\vec R, 0, E) \vec S_1\cdot\vec s\, \big] \  dE,
\label{final}
\end{align}
We left out here in Eq. (\ref{final}), the term involving $G^A$, which turns out to be zero, i. e., 
\begin{align}
I = & \frac{ \lambda^2}{\pi} \ {\rm Im} \int_{-\infty}^{E_F}  {\rm Tr} \big[G^A(0,\vec R, E)\, \vec S_2\cdot\vec s \nonumber \\
& G(\vec R, 0, E) \vec S_1\cdot\vec s\, \big] \  dE  = 0.
\end{align}
This can be easily shown by expanding the Green's functions in terms of the Pauli matrices 
\begin{align}
&G(\vec R, 0, E)  = g_0 \sigma_0 + \sum_{i=1}^3 g_i \sigma_i,   \\ \nonumber
&G^A(0, \vec R, E)  = g_0^* \sigma_0 + \sum_{i=1}^3 g_i^* \sigma_i,
\end{align}
using the result  ${\rm Tr}  \ (A + B) = {\rm Tr} \ A + {\rm Tr} \ B$  and the  trace equalities
\begin{align}
&{\rm Tr} \ (\sigma_i\sigma_j)= 2\delta_{ij},     \\ \nonumber
&{\rm Tr} \  (\sigma_i\sigma_j\sigma_k)= 2i \ \varepsilon_{ijk},   \\ \nonumber
&{\rm Tr} \ (\sigma_i\sigma_j\sigma_k\sigma_l)=2(\delta_{ij}\delta_{kl}-\delta_{ik}\delta_{jl}+\delta_{il}\delta_{jk}),
\end{align}
written in terms of the Kronecker deltas $\delta_{ij}$ and the Levi-Civita symbols $\varepsilon_{ijk}$.

\end{document}